\documentclass[journal,twocolumn]{IEEEtran}

\usepackage{amsmath,amsthm,amssymb,scrextend}
\usepackage{fancyhdr}
\usepackage{relsize}
\usepackage{epstopdf} 
\usepackage{floatrow}
\usepackage[font={small}]{caption}
\usepackage{multirow}
\usepackage{balance}
\usepackage[numbers,sort&compress,square]{natbib}
\usepackage{subfigure}
\usepackage{ifpdf}
\ifCLASSINFOpdf
   \usepackage[pdftex]{graphicx}
   \graphicspath{{./Figures/pdf/}}
   \DeclareGraphicsExtensions{.pdf}
\else
   \usepackage[dvips]{graphicx}
   \graphicspath{{./Figures/eps/}}
   \DeclareGraphicsExtensions{.eps}
\fi

\begin{document}

\pagenumbering{gobble} 

\title{The Impact of Correlated Blocking on Millimeter-Wave Personal Networks}
\author{
\IEEEauthorblockN{Enass Hriba and 
Matthew C. Valenti
 } \\
West Virginia University, Morgantown, WV, USA. \\
\vspace{-0.5cm}
}

\maketitle
\thispagestyle{empty}


\begin{abstract}
Due to its potential to support high data rates at low latency with reasonable interference isolation, millimeter-wave (mmWave) communications has emerged as a promising solution for wireless personal-area networks (WPAN) and an enabler for emerging applications such as high-resolution untethered virtual reality. At mmWave, signals are prone to blockage by objects in the environment, including human bodies. Most mmWave systems utilize directional antennas in order to overcome the significant path loss. In this paper, we consider the effects of blockage and antenna directivity on the performance of a mmWave WPAN. Similar to related work, we assume that the interferers are in arbitrary locations and the blockages are drawn from a random point process. However, unlike related work that assumes independent blocking, we carefully account for the possibility of correlated blocking, which arises when two interferers are close to each other and therefore an obstruction that blocks the first interferer may likely block the second interferer. Closed form expressions for the blockage correlation coefficient and the distribution of the SINR are provided for the case of two dominant interferers and a fixed number of blockages drawn from a binomial point process. Finally, the effects of antenna directivity and the spatial randomness of the interferers are taken into account, resulting in SINR curves that fully account for correlated blocking, which are compared against curves that neglect correlation. The results provide insight into the validity of the commonly held assumption of independent blocking and the improved accuracy that can be obtained when the blocking correlation is taken into account.

\end{abstract}


\vspace{0.05cm}

\section{Introduction}

Communicating at millimeter-wave (mmWave) frequencies is attractive due to the potential to support high data rates at low latency \cite{Rapp:2013,Akdeniz}.  The mmWave band is characterized by high attenuation, which is both a blessing and a curse \cite{rappaport2014millimeter}.  On the one hand, the desired signal is highly attenuated, and to overcome the attenuation, high gain directional antennas are required;  however, due to the small wavelength, compact multi-element antenna arrays are feasible, even on a compact user terminal.   On the other hand, interference tends to also be highly attenuated, and thus the band is characterized as having reasonable interference isolation.

Due to these characteristics, mmWave has emerged as a promising solution for wireless personal-area networks (WPAN) and as an enabler for emerging applications such as high-resolution untethered virtual reality, augmented reality, and mixed reality \cite{Fisher2007,Singh2009,YNiu2015}.  These technologies have significant military applications, as virtual reality can help to better train the warfighter, while mixed/augmented reality has the potential to provide enhanced situational awareness.

Another characteristic of mmWave is that it is prone to blockage by objects in the environment, including human bodies.  On the battlefield, blockages may include soldiers, tanks, helicopters, and other equipment creating a dynamic environment characterized by changing blocking conditions.  Blocking makes it especially difficult to provide universal coverage with a cellular infrastructure.  For instance, blockage by walls provides isolation between indoor and outdoor environments, making it difficult for an outdoor base station to provide coverage indoors \cite{BaiVazeHeath}.

The performance of mmWave systems can be characterized by the outage probability, or equivalently, by the cumulative distribution function (CDF) of the signal-to-interference ratio (SINR). Alternatively, the performance can be characterized by the coverage probability, which is the complimentary CDF of the SINR, or the rate distribution, which can be found by using information theory to link the SINR to the achievable rate.  
Prior work has considered the SINR distribution of mmWave personal networks \cite{Venugopal2015,Venugopal2016,Hriba2017}.  Such work assumes that the blockages are drawn from a point process (or, more specifically, that the centers of the blockages are drawn from a point process and each blockage is characterized by either a constant or random width).  Meanwhile, the interferers are either in fixed locations or their locations are also drawn from a point process.  A universal assumption in this prior art is that the blocking is independent; i.e., each interferer is blocked independently from the other interferers.

In actual networks, blocking might be correlated.  This is particularly true when two interferers are close to each other, or more generally when two interferers have a similar angle to the reference receiver.  In this case, when one interferer is blocked, there is a significant probability that the other interferer is also blocked.  However, correlated blocking can arise even when interferers are not close.  Take, for instance, an extreme case where there is just one blockage in the environment and two interferers located far apart from each other.  If the first interferer is blocked, then the second one cannot be blocked, giving rise to a \emph{negative} correlation.

As blocking has a major influence on the distribution of the interference, it must be carefully taken into account.  Independent blocking is a crude approximation that fails to accurately capture the true environment, especially when the interferers are closely spaced or when there are few sources of blocking.   We note that blocking can be correlated even when the sources of blockage are placed independently according to a point process.  The issue of blockage correlation was recently considered in \cite{Aditya17,Aditya18}, but it was in the context of a localization application where the goal was to ensure that a minimum number of positioning transmitters were visible by the receiver.  As such, it was only concerned with the \emph{number} of unblocked transmissions rather than the distribution of the received aggregate signal (i.e., the interference power).

In this paper, we accurately characterize the performance of mmWave WPAN systems in the presence of correlated blocking.   We assume that an arbitrary number of blockages are placed according to a point process.  For ease of exposition, we consider the case of two dominant interferers, though the methodology can be extended to multiple interferers.  The signal model is such that blocked signals are completely attenuated, while line-of-site (LOS), i.e., non-blocked, signals are subject to an exponential path loss and additive white Gaussian noise (AWGN).  Though it complicates the exposition and notation, the methodology can be extended to more elaborate models, such as one wherein all signals are subject to fading and NLOS signals are only partially attenuated (see, e.g., \cite{Hriba2017}).

The remainder of the paper is organized as follows.  We begin by providing the system model in Section II, wherein there are two interferers and an arbitrary number of blockages, each drawn from a binomial point process.  Section III provides an analysis of the SINR distribution, where the results depend on the blockage correlation coefficient.  Section IV derives the blockage correlation coefficient using arguments based on the geometry and the properties of the blockage point process; i.e., by using stochastic geometry.  Section V considers furthermore the effects of antenna directivity and randomly placed interferers by allowing the interfering transmitters to have a random location and orientation.  The section leverages the analysis provided in \cite{Valenti2014}.  Finally, Section VI concludes the paper, suggesting extensions and generalizations of the work.

\section{System Model}
Consider a mmWave WPAN consisting of a reference transmitter-receiver pair surrounded by both blockages and interfering transmitters.  The network is contained in an arbitrarily shaped region $A$, where the variable $A$ is used to denote both the region and its area.  Any additional sources of interference located outside of $A$ are assumed to be completely attenuated and therefore do not directly factor into the performance of the network, though they may contribute to the noise floor.

The goal of our work is to investigate the influence of correlated blocking on the system performance.   While it is possible for there to be more than two sources of interference, and these interferers may be subject to correlated blocking, the main concept is best exposed by limiting the discussion to just two interferers.  As mmWave systems tend to be limited by a few dominant interferers, this limitation is a practical one in most cases.  However, Section VI contemplates ways to extend the analysis to the case of multiple interferers.  

\begin{figure}[t]
\centering
\includegraphics[scale=0.32]{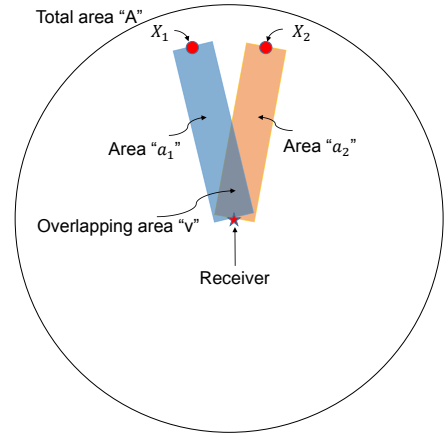}
\vspace{-0.25cm}
\caption{Example network topology.  Denoted by a red star, the receiver is located at the center of the circular region $A$.  Denoted by red dots, the two interferers are in the northern part of $A$.  The blocking zones of each interferer are indicated by colored rectangles, and their intersection is represented by $v$.\vspace{-0.6cm}}
\label{fig:OverlappingArea}
\end{figure}

Hence, there are three transmitters: A source transmitter and two interferers.  Let the variable $X_i$ denote the transmitter and its location.  In particular, let $X_0$ denote the source transmitter, and $X_1$ and $X_2$ denote the interferers.  Each location is represented by a complex number $X_i=R_ie^{j\phi_i}$, where $R_i$ represents the distance from the receiver to the $i^{th}$ transmitter and $\phi_i$ represents the (azimuth) angle from the receiver to the transmitter $X_i$.  Without loss of generality, $R_1 \leqslant R_2$.  Fig. \ref{fig:OverlappingArea} shows an example of the network topology.  Here, the network region $A$ is a circle, though our methodology does not require $A$ to be any particular shape.  The receiver is at the center of $A$ (indicted by the red star), and the two interferers are located in the northern part of $A$ (indicated by the red dots).

Within the network, there are $K$ distributed blockages. As in \cite{Aditya18}, each blockage is modeled as a point, a line segment of length $W$ centered at the point, and the line segment's orientation angle.   The points are distributed according to a binomial point process, and as such, they are independently and uniformly placed on $A$.   As in \cite{Aditya18}, the orientation angles are selected such that the line segment is perpendicular to the line between the receiver and the center of the blockage, which is equivalent to saying that the line segment is actually the projection of the visible face of the blockage rather than the entire object.   Although $W$ can itself be random, we assume here that all blockages have the same value of $W$. If a line segment cuts the path between $X_i$ and the receiver, then the signal from $X_i$ is non line-of-sight (NLOS), while otherwise it  is line of sight (LOS).  Here, we assume that NLOS signals are completely blocked while LOS signals experience exponential path-loss with a path-loss exponent denoted by $\alpha$.

Each interferer has a \emph{blockage region} associated with it, indicated by the colored rectangles in Fig. \ref{fig:OverlappingArea}.  We use $a_i$ to denote the blockage region associated with $X_i, i = \{1,2\}$, and  its area.  If the center of a blockage falls within $a_i$, then $X_i$ will be blocked since at least some part of the blockage will intersect the path between $X_i$ and the receiver.    From the given geometry, it is clear that $a_i = W R_i$.  Unless $X_1$ and $X_2$ are exactly on opposite sides of $A$, i.e. $|\phi_1-\phi_2|=\pi$, there will be an overlapping region $v$ common to both $a_1$ and $a_2$.  Because of the overlap, it is possible for a single blockage to block both $X_1$ and $X_2$ if the blockage falls within region $v$.  This is an example of positive blockage correlation.  However, it is also possible to have negative blockage correlation.  For instance, if there is just a single blockage (i.e., $K=1$), then if the blockage lies in region $a_1 \setminus v$ (i.e., in $a_1$ but not in $v$), then it cannot be in $a_2$.  In this case, $X_1$ will be blocked, but $X_2$ cannot be blocked.  As we will show, negative correlation is also possible even when $K>1$.





\section{SINR Outage Analysis} 
 The signal-to-interference and noise ratio (SINR) at the receiver is given by 
\begin{eqnarray}
   \mathsf{SINR} = \frac{\Omega_0}{c+\displaystyle \sum_{i=1}^2 (1-B_i)\Omega_i}
   \label{eg:S}
\end{eqnarray}
where ${\Omega_i}=R_i^{-\alpha}$  is the received power from transmitter $X_i$ (recalling that $X_0$ is the source transmitter, $X_1$ and $X_2$ are the two interferers, and $\alpha$ is the path-loss exponent),  the constant $c$ is selected so that the signal-to-noise ratio (SNR) is the value of SINR when the interference is turned off ($\mathsf{SNR}=\Omega_0/c \rightarrow c = \Omega_0/\mathsf{SNR}$), and $\{ B_1, B_2 \}$ are a pair of Bernoulli random variables, which may in general be correlated.  

The variable $B_i$ is used to indicate that $X_i$ is blocked, and thus when $B_i = 1$, $\Omega_i$ does not factor into the SINR.  Let $p_{B_1,B_2}(b_1,b_2)$ be the joint probability mass function (pmf) of $\{ B_1, B_2 \}$. Let $p_i$ denote the probability that $B_i = 1$.  Furthermore, let $q_i = 1 - p_i$. 
Letting $\rho$ denote the correlation coefficient between $B_1$ and $B_2$, the joint pmf is found to be
\begin{eqnarray}
  p_{B_1,B_2}(b_1,b_2)
  \hspace{-0.2cm}& = &\hspace{-0.2cm}
  \begin{cases}
  q_1 q_2  +\rho h &  \mbox{for $b_1 = 0, b_2 = 0$} \\  
  q_1 p_2 - \rho h &  \mbox{for $b_1 = 0, b_2 = 1$} \\   
  p_1q_2 - \rho h  &  \mbox{for $b_1 = 1, b_2 = 0$} \\  
  p_1 p_2 + \rho h  &  \mbox{for $b_1 = 1, b_2 = 1$} \\    
 \end{cases}
 \label{eq:pB1B2}
\end{eqnarray}
where $h = \sqrt{p_1 p_2 q_1 q_2}$. A proof of (\ref{eq:pB1B2}) can be found in the Appendix.  See also \cite{Fontana:2017}. 



Because blockages are uniformly distributed over $A$, the probability that a given blockage lands in $a_i$ is equal to $a_i/A$, and hence the probability it is outside $a_i$ is $1-a_i/A$. Since blockages are independently placed, the probability that all $K$ blockages are outside $a_i$ is $(1-a_i/A)^K$, and when this occurs, $X_i$ will be be LOS (i.e., not blocked).  Conversely, $X_i$ will be NLOS (i.e., blocked) when not all of the blockages are outside $a_i$, which occurs with probability
\begin{eqnarray}
p_i=1-\left(1-\frac{a_i}{A}\right)^K. \label{eq:pi}
\end{eqnarray}


The goal of this section to formulate the CDF of the SINR 
$F_\mathsf{SINR}(\beta)$, which quantifies the likelihood that the SINR at the receiver is below some threshold $\beta$.  If $\beta$ is interpreted as the minimum acceptable SINR required to achieve reliable communications, then $F_\mathsf{SINR}(\beta)$ is the \emph{outage probability} of the system $P_o(\beta) = F_\mathsf{SINR}(\beta)$.  The \emph{coverage probability} is the \emph{complimentary} CDF, $P_c(\beta) = 1-F_\mathsf{SINR}(\beta)$ and is the likelihood that the SINR is sufficiently high to provide coverage.  The \emph{rate} distribution can be found by linking the threshold $\beta$ to the transmission rate, for instance by using the appropriate expression for channel capacity.

The CDF of SINR evaluated at threshold $\beta$ can be determined as follows:
\begin{eqnarray}
F_\mathsf{SINR}(\beta)&=&P\left[\mathsf{SINR}\leq \beta \right]\nonumber \\
&=&1-P\bigg[\underbrace{\sum_{i=1}^2 \Omega_i(1-B_i)}_{Z} \leq \frac{\Omega_0}{\beta}-c\bigg]\nonumber \\
&=&1-F_{Z}\left(\frac{\Omega_0}{\beta}-c\right).\label{cdfsinr}
\end{eqnarray}
The discrete variable $Z$ represents the sum of the unblocked interference. To find the CDF of $Z$ we need to find the probability of each value of $Z$, which is found as follows.
The probability that $Z=0$ can be found by noting that $Z=0$ when both $X_1$ and $X_2$ are blocked. From (\ref{eq:pB1B2}), this  is
\begin{eqnarray}
p_Z(0)=p_{B_1,B_2}(1,1)=p_1p_2+\rho h. \label{eq:Z=0}
 \end{eqnarray}
The probability that $Z=\Omega_i , i \in \{1,2\}$ can be found by noting that $Z=\Omega_i$ when only $X_i$ is LOS. From (\ref{eq:pB1B2}), this  is
 \begin{eqnarray}
p_Z(\Omega_1)&=&p_{B_1,B_2}(0,1)=q_1 p_2-\rho h. \label{eq:Z=1} \\
p_Z(\Omega_2)&=&p_{B_1,B_2}(1,0)=p_1 q_2-\rho h. \label{eq:Z=2}
 \end{eqnarray}
Finally, by noting that $Z=\Omega_1+\Omega_2$ when both $X_1$ and $X_2$ are LOS leads to
\begin{eqnarray}
p_Z({\Omega_1+\Omega_2})&=&p_{B_1,B_2}(0,0)=q_1 q_2+\rho h. \label{eq:Z=1+2}
\end{eqnarray}

\begin{figure}[t]
\centering
    \includegraphics[width=\textwidth]{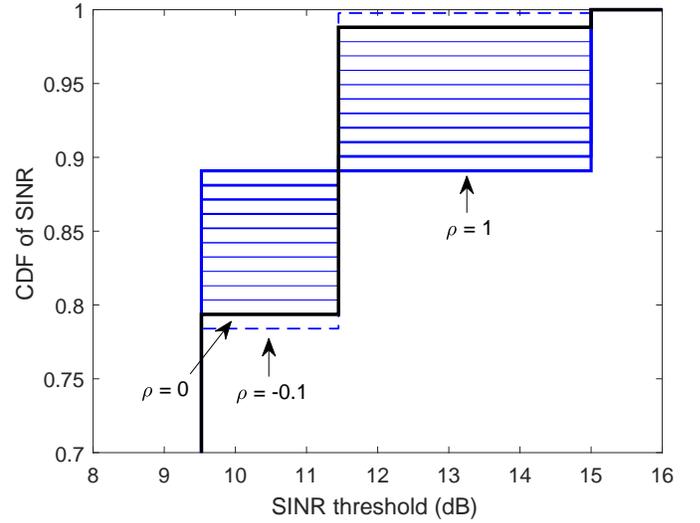}
    \vspace{-0.5cm}
\caption{The CDF of the SINR $F_\mathsf{SINR}(\beta)$ for different values of $\rho$. The thick black line shows the CDF when $\rho=0$, the dashed blue line shows the case when $\rho=-0.1$, and the solid blue lines correspond to positive values of $\rho$ in increments of 0.1.\vspace{-0.5cm}}
\label{fig:rho}
\end{figure}

%





From (\ref{eq:Z=0}) to (\ref{eq:Z=1+2}), the CDF of $Z$ is found to be:
 \begin{eqnarray}
F_{Z}\hspace{-0.1cm}\left(z\right)\hspace{-0.1cm}=\hspace{-0.1cm}\begin{cases}
    0       & \smallskip \hspace{-0.2cm} \text{for} \enskip  z<0 \\
   p_1p_2 + \rho h & \smallskip \hspace{-0.2cm}\text{for} \enskip  0\leq z <\Omega_2 \\
    p_1  & \smallskip \hspace{-0.2cm}\text{for} \enskip \Omega_2 \leq z<\Omega_1 \\
     p_1 +q_1p_2-\rho h & \smallskip \hspace{-0.2cm}\text{for} \enskip \Omega_1 \leq z<  \Omega_1+\Omega_2  \\  
    1 & \smallskip \hspace{-0.2cm}\text{for} \enskip z \geq  \Omega_1+\Omega_2.  
  \end{cases}\hspace{-.3cm}
\end{eqnarray}

When $R_1=R_2$, and thus $\Omega_1=\Omega_2=\Omega$, $p_1=p_2=p$, and $q_1=q_2=q$, the CDF is
 \begin{eqnarray}
F_{Z}\left(z\right)=\begin{cases}
    0       & \smallskip  \text{for} \enskip  z<0 \\
   p^2 + \rho pq & \smallskip \text{for} \enskip  0\leq z <\Omega \\
     1-\rho pq-q^2 & \smallskip \text{for} \enskip \Omega \leq z<  2\Omega  \\  
    1 & \smallskip \text{for} \enskip z \geq  2\Omega.
  \end{cases}
  \label{eq:CDF}
\end{eqnarray}


Fig. \ref{fig:rho} shows the effect that the value of the correlation coefficient $\rho$ has upon the CDF of SINR, which is found by substituting (\ref{eq:CDF}) into (\ref{cdfsinr}). The curves were computed with $R_1=R_2=5$, $\alpha=2$, and $\mathsf{SNR} = 15$ dB. The value of $p$ was computed using (\ref{eq:pi}) by assuming $W=1$, $K=20$, and that $A$ is a circle of radius 6.  The CDF is found assuming values of $\rho$ between $\rho=-0.1$ to $\rho=1$ in increments of 0.1.  The thick black line represents the case that $\rho=0$, corresponding to uncorrelated blocking. The dashed blue line represents the case when $\rho=-0.1$. The solid blue lines correspond to positive values of $\rho$ in increments of 0.1, where the thinnest line corresponds to $\rho=0.1$ and the thickest line corresponds to $\rho=1$.

Fig. \ref{fig:rho} shows a first step up at 9.5 dB, and the increment of the step is equal to the probability that both interferers are LOS.   The magnitude of the step gets larger as the blocking is more correlated, because (positive) correlation increases the chance that both interferers are  LOS (i.e., $p_{B_1,B_2}(0,0)$).  Negative correlation actually reduces the magnitude of the step.  The next step up occurs at 11.5 dB, which is the SINR when just one of the two interferers is blocked.  The magnitude of this jump is equal to the probability that just one interferer is blocked, and this magnitude decreases with positive correlation.  Finally, there is a step at 15 dB, which corresponds to the case that both interferers are blocked, in which case the SINR equals the SNR.  Notice that when $\rho=1$, the two middle steps merge.  This is because when $\rho=1$, it is impossible for just one interferer to be blocked, so the curve goes directly from SINR = 9.5 dB to SINR = 15 dB.
\vspace{-0.1cm}


\section{Blockage Correlation Coefficient}

Let's now consider how to find $\rho$, the blockage correlation coefficient.  From (\ref{eq:pB1B2}), 
\begin{eqnarray}
  \rho
  & = &
  \frac{ p_{B_1,B_2}(0,0) - q_1 q_2 }{ h }
\end{eqnarray}
where $p_{B_1,B_2}(0,0)$ is the probability that both $X_1$ and $X_2$ are not blocked. Looking at Fig. \ref{fig:OverlappingArea}, this can occur when all blockages are outside areas $a_1$ and $a_2$. Taking into account the overlap $v$, this probability is 
\vspace{-0.1cm}
\begin{eqnarray}
 p_{B_1,B_2}(0,0)= \left( 1-\frac{a_1+a_2-v}{A} \right)^K
\end{eqnarray}

\begin{figure}[t]
\centering
    \includegraphics[width=0.9\textwidth]{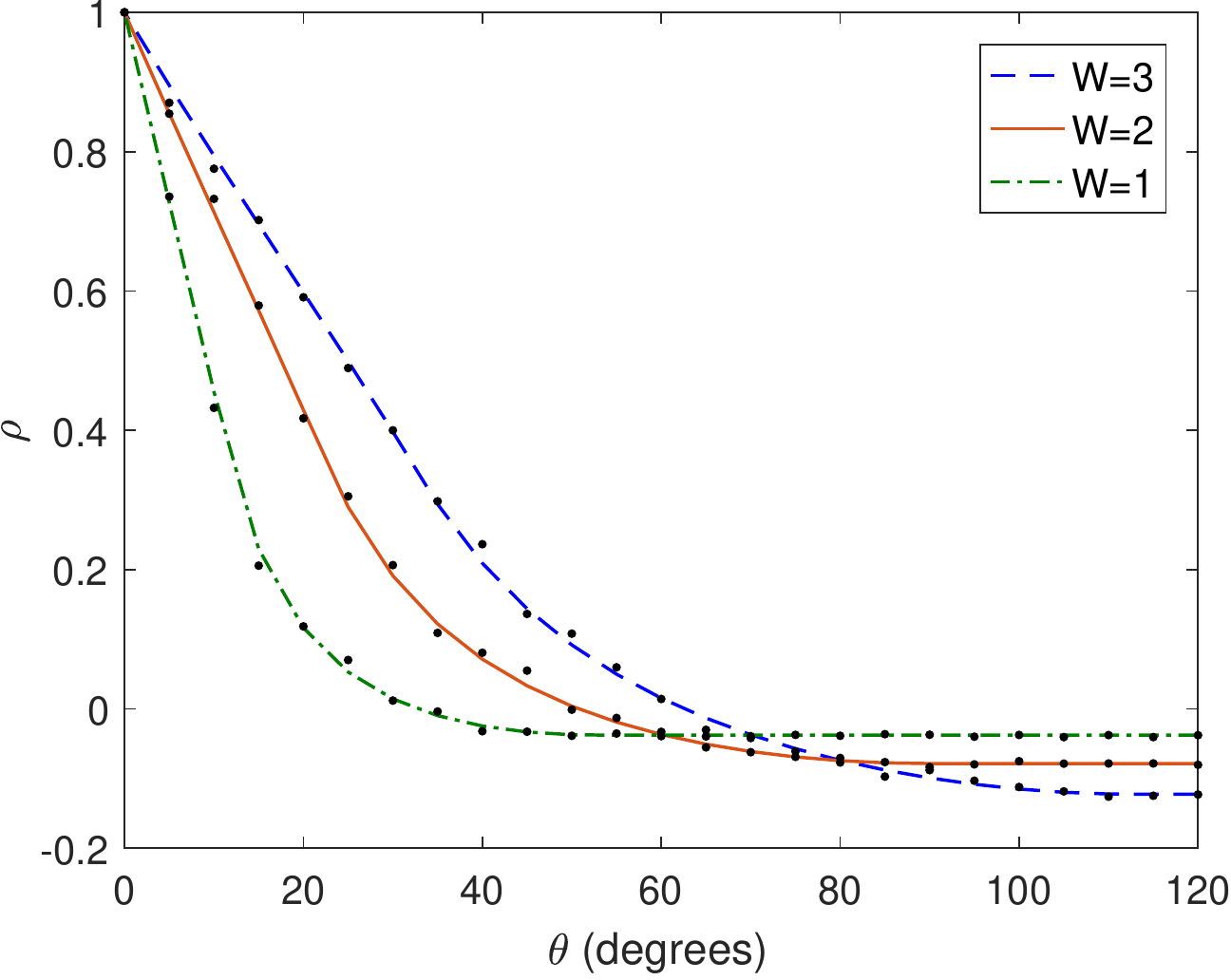}
    \vspace{-0.1cm}
\caption{The correlation coefficient ($\rho$) versus the angular separation between $X_1$ and $X_2$ ($\theta=\hspace{0.1cm}\mid\phi_1-\phi_2\mid$) for different values of blockage width ($W$). The black dots represent simulation results. The curves show the exact analytical expression found using the methods of this paper.    \vspace{-0.1cm}}
\label{fig:rho_vs_theta_W}
\end{figure}
  
  \begin{figure}[t]
\centering
    \includegraphics[width=0.9\textwidth]{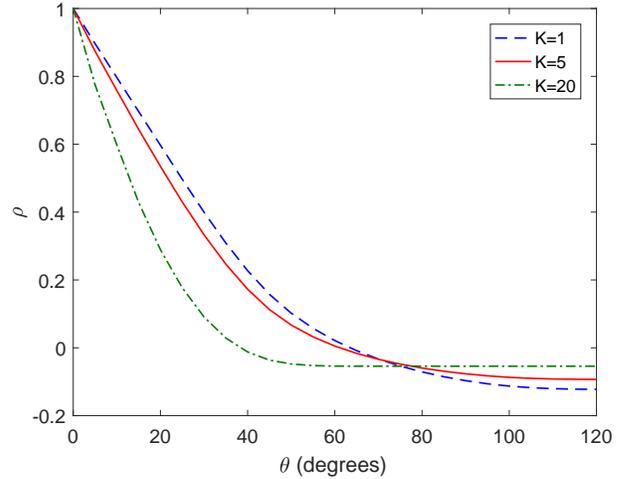}
    \vspace{-0.1cm}
\caption{The correlation coefficient ($\rho$) versus the angular separation between $X_1$ and $X_2$ ($\theta=\hspace{0.1cm}\mid\phi_1-\phi_2\mid$) for different values of number of blockages ($K$). 
\vspace{-0.5cm}}
\label{fig:rho_vs_theta_K}
\end{figure}

\begin{figure}[t]
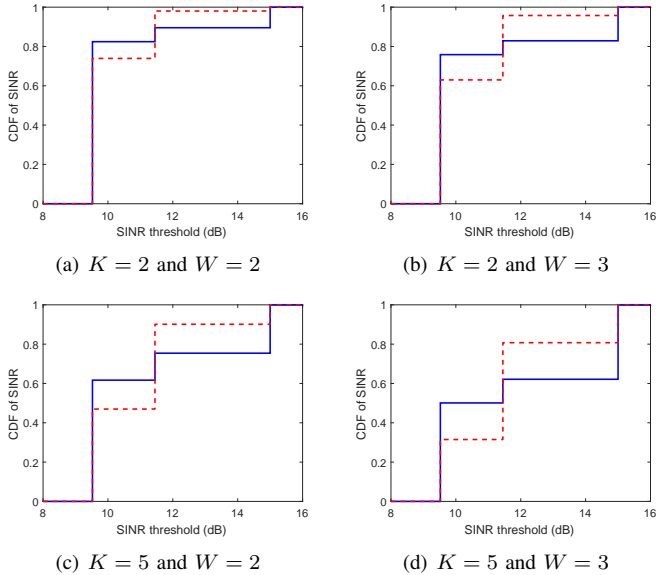

\centering
\begin{subfigure}[$K=2$ and $W=2$]{
\includegraphics[width=.45\linewidth] {K2W2-crop}   
\label{fig:subfig1} }
\end{subfigure}
\hfill
\begin{subfigure}[$K=2$ and $W=3$]{
\includegraphics[width=.45\linewidth] {K2W3-crop}
\label{fig:subfig2} }%
\end{subfigure}
\hfill
\begin{subfigure}[$K=5$ and $W=2$]{
\includegraphics[width=.45\linewidth] {k5w2-crop}
\label{fig:subfig3} }%
\end{subfigure}%
\hfill
\begin{subfigure}[$K=5$ and $W=3$]{
\includegraphics[width=.45\linewidth] {K5W3-crop}
\label{fig:subfig4} }%
\end{subfigure}%
 
\caption{Comparison of CDF of SINR that accounts for correlated blockage (solid blue line) against independent blockage (dotted red line). 
\vspace{-0.5cm}}
\label{fig:KvsW}
\end{figure}

Fig. \ref{fig:rho_vs_theta_W} and Fig. \ref{fig:rho_vs_theta_K} show $\rho$ as a function of the angular separation ($\theta$) between $X_1$ and $X_2$,  where $\theta=\hspace{0.1cm}\mid\phi_1-\phi_2\mid$. As with Fig. \ref{fig:rho}, $R_1=R_2=5$, and $A=2\pi6^2$. In Fig. \ref{fig:rho_vs_theta_W}, a fixed value of $K=1$ is used and $W$ is varied. In Fig. \ref{fig:rho_vs_theta_K}, $W=3$ and the value of $K$ is varied. In Fig. \ref{fig:rho_vs_theta_W}, the black dots represents simulation results, which are shown merely to confirm the validity of the approach.

Both figures show that $\rho$ decreases with increasing $\theta$.  This is because the area $v$ gets smaller as $\theta$ increases.  As $\theta$ approaches $\pi$ radians, $v$ approaches zero, and the correlation is minimized.  Note that the minimum value is actually less than zero, showing the possibility of negative correlation.   The negative correlation can occur when $a_1$ and $a_2$ are non-overlapping because if $X_1$ is blocked by $m$ blockages, then there are only $K-m$ blockages left that could possibly block $X_2$.  The figures show that correlation is more dramatic when $W$ is large, since a single large blockage is likely to simultaneously block both interferers, and when $K$ is small, which corresponds to the case that there are fewer blockages.

Fig. \ref{fig:KvsW} shows the CDF of the SINR for the same network with different values of $W$ and $K$.  The area $A$ is again a circle with radius 6, $R_1=R_2=5$, SNR = 15 dB and $\theta= 25^\circ$. The solid blue line shows the SINR distribution found by accurately accounting for correlated blocking; i.e., by using the methodology of this section to find $\rho$.  The dashed red line corresponds to the CDF assuming blocking is independent; i.e., fixing $\rho=0$. From (\ref{eq:CDF}), the difference between the two curves is equal to $\rho p q$, and thus the difference will grow if $\rho$ increases and/or $p$ increases, as long as $p < 0.5$.  By comparing Fig. \ref{fig:subfig1} to Fig. \ref{fig:subfig2} or Fig. \ref{fig:subfig3} to Fig. \ref{fig:subfig4}, the gap between the two CDFs increases as $W$ goes from 2 to 3. This can be explained by referring to Fig. \ref{fig:rho_vs_theta_W}, which shows that $\rho$ increases for sufficiently small $\theta$ as $W$ increases.  By comparing Fig. \ref{fig:subfig1} to Fig. \ref{fig:subfig3} or Fig. \ref{fig:subfig2} to Fig. \ref{fig:subfig4}, the gap between the two CDFs increases as $K$ goes from 2 to 5. Although  Fig. \ref{fig:rho_vs_theta_K} shows that $\rho$ decreases with increasing $K$ at moderate $\theta$, this behavior is offset by the fact that $p$ increases with $K$, per (\ref{eq:pi}).  Thus the gap actually increases with increasing $K$.

\vspace{-0.1cm}
\section{Antenna directivity and spatial randomness}
\begin{figure}[t]
\centering
    \includegraphics[width=\textwidth]{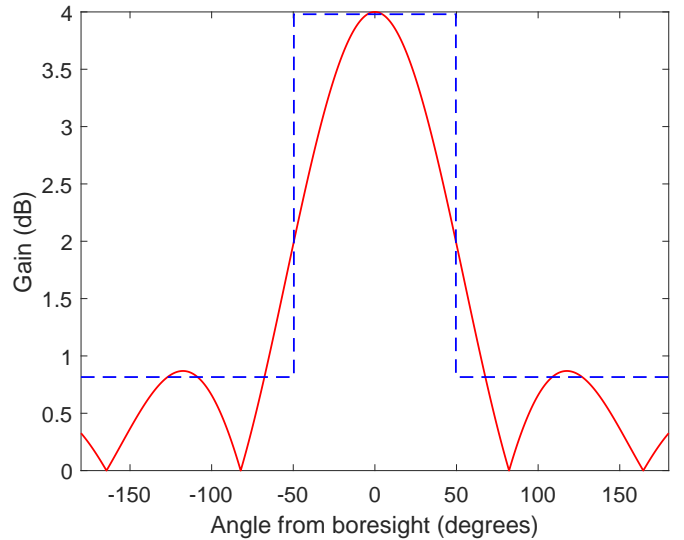}
\caption{Actual antenna model (solid red line) versus sectorized antenna model (dashed blue line).\vspace{-0.5cm}}
\label{fig:AntennaGainPattern}
\end{figure}

Thus far, we have assumed that the interferers are in fixed locations and the antennas are omnidirectional. In practice, the locations of the interferers may themselves be random, and directional antennas may be used. When the antennas are directional, the received power from the $i^{th}$ interferer is 
\begin{eqnarray}
\Omega_i
& = & 
g_r(\phi_i)g_t( |\phi_i - \psi_i| - \pi)R_i^{-\alpha}
\end{eqnarray}
where $g_r(.)$ is the antenna gain of the receiver, $g_t(.)$ is the antenna gain of the transmitter and $\psi_i$ is the azimuth angle of interferer's transmit antenna. Here, we assume the antenna patterns are function of the azimuth angle only. The red solid line in Fig. \ref{fig:AntennaGainPattern} shows an example antenna pattern for a 4-element planar array \cite{Anderson2009}.

Often, in the mmWave literature, the exact antenna pattern is approximated with a sectorized model, where the antenna gain is just one of two values corresponding to the main lobe and the side lobe. The blue dashed line in Fig. \ref{fig:AntennaGainPattern} shows a sectorized antenna that approximates the actual antenna pattern \cite{Venugopal2015,Venugopal2016}.

The direction $\psi_i$ that $X_i$ is pointing is generally unknown and may be modeled as a random variable. Since $\Omega_i$ is a function of $\psi_i$, it is then a random variable, even if $X_i$ is in a fixed location. For a WPAN, we may assume $\psi_i$ is uniformly distributed from 0 to $2\pi$ radians; i.e, $\psi_i \sim U(0,2\pi).$

\begin{figure}[t]
\centering
    \includegraphics[width=\textwidth]{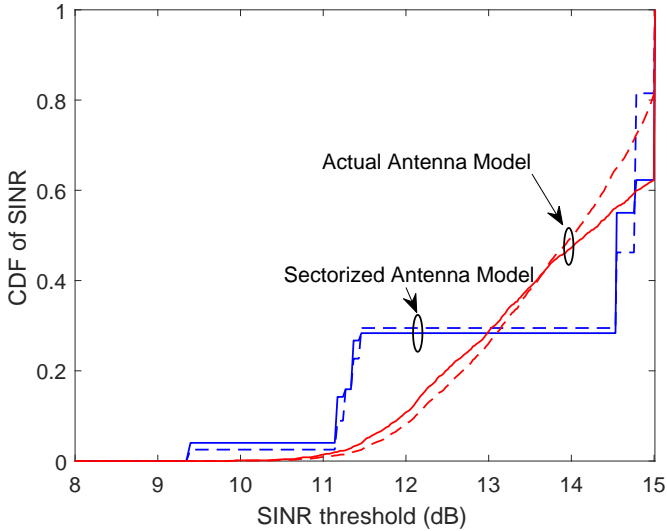}
\caption{Comparison of the CDF of SINR for a particular network realization with the actual antenna pattern (red curves) vs. with a sectorized antenna model (blue curves).  Solid lines account for correlated blocking while dashed lines assume independent blocking. }
\label{fig:CorrelatedIndependentBlocking_SectorizedActualPattern}
\end{figure}

Fig. \ref{fig:CorrelatedIndependentBlocking_SectorizedActualPattern} shows performance when antenna directivity is taken into account.  Two pairs of curves are shown.  The first pair (in red) shows the CDF when the actual antenna pattern of Fig. \ref{fig:AntennaGainPattern} is used.  The second pair (in blue) shows the CDF when the sectorized model of Fig. \ref{fig:AntennaGainPattern} is used as an approximation.  For each pair of curves, one curve (with a solid line) shows the CDF when the blockage correlation is taken into account and the other curve (with the dashed line) shows the CDF when the blocking is assumed to be independent. The curves are generated by again assuming that $A$ is a circle of radius 6 and an SNR = 15 dB.  There are $K=5$ blockages of width $W=2$.  The interferers are in fixed locations with   $R_1=4$, $R_2=5$, and $\theta=25^\circ$.  We note that there is a significant difference in this case between the CDFs predicted using the actual antenna model vs. the curves generated using the sectorized approximation.  To a lesser extent, there is a difference for each antenna model between the curve that accounts for correlation and the one that assumes independent blocking, and the difference is more pronounced at higher SINR thresholds.

\begin{figure}[t]
\centering
    \includegraphics[width=\textwidth]{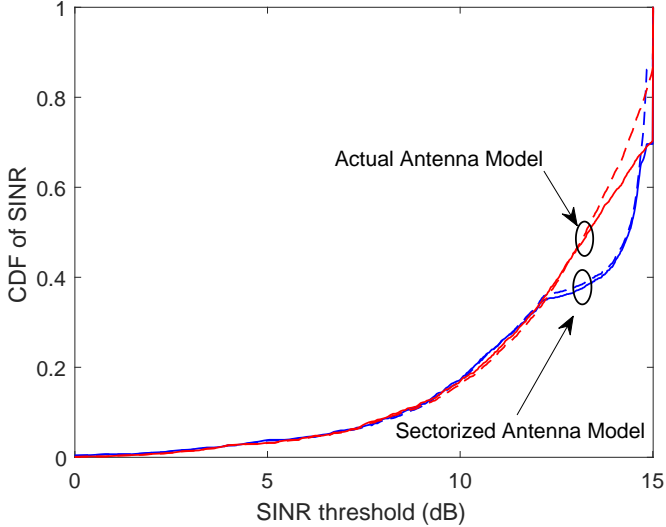}
\caption{The CDF of SINR when averaged over 1000 network network realizations. 
Red curves use an actual antenna pattern and blue curves use a sectorized antenna model. Solid lines account for correlated blocking while dashed lines assume independent blocking. \vspace{-0.6cm}}
\label{fig:RandomSectorizedDirectivity_IndependentVsCorrelationBlockage}
\end{figure}

If, in addition, the location of the $X_i$ are random, then each $\Omega_i$ is a random variable that depends on both the location and directivity of the interferers. Fig. \ref{fig:RandomSectorizedDirectivity_IndependentVsCorrelationBlockage} uses the same parameters that were used in Fig. \ref{fig:CorrelatedIndependentBlocking_SectorizedActualPattern} except that now the interferers are randomly placed.  In particular, the two interferers are placed independently and uniformly within the circular area $A$.  The CDF is found by averaging over 1000 such placements (i.e., network realizations).  When the locations are random, the difference between the dashed and solid curves begins to tighten up, implying that the effect of correlation is less important.  This is because when randomly placed, the two interferers are often far apart from one another.  However, correlation is important for certain regions of the plot, particularly at high values of threshold. Moreover, the difference between the two antenna models is less pronounced, especially at lower values of SINR threshold.

\section{Conclusion}

In a mmWave WPAN system, the interference may be subject to correlated blocking.  This is true even if the individual blockages are independently placed, since it is possible for a single blockage to block multiple interferers if the blockage is sufficiently wide and the interferers sufficiently close.  The commonly held assumption of independent blocking leads to an incorrect characterization of the performance of the system, for instance, when it is quantified by the distribution of the SINR.  By using the methodology in this paper, the correlation between two sources of interference may be found and factored into the analysis, yielding more accurate results.

The analysis can be extended in a variety of ways.  In Section V, we have already shown that it can be combined with an analysis that accounts for antenna directivity and the random location and orientation of the interferers.   While this paper has focused on the extreme case that LOS signals are AWGN while NLOS signals are completely blocked, it is possible to adapt the analysis to more sophisticated channels, such as those where both LOS and NLOS signals are subject to fading and path loss, but the fading and path loss parameters are different depending on the blocking state.  See, for instance, \cite{Hriba2017} for more detail.

Finally, while this paper focused on the pairwise correlation between two interferers, it can be extended to the more general case of an arbitrary number of interferers.  One way to handle this is to only consider the correlation of the two closest interferers (the most dominant ones) while assuming that all other interferers are subject to independent blocking.  Another solution is to group interferers into pairs, and only consider the pairwise correlation, while neglecting higher order effects.  As performance in a mmWave system is typically dominated by just a few interferers, we anticipate that either of these approaches would yield accurate results. We also anticipate that when several interferers are present, the effects of correlation will be even more pronounced in a random network, as the likelihood that two interferers are close together increases with the number of interferers.
\section*{Appendix}

The correlation coefficient between $B_1$ and $B_2$ is given by
\begin{eqnarray} 
 \rho&=&\frac{E[B_1{B_2}]-E[B_1]E[{B_2}]}{\sqrt{\sigma_{B_1}^2\sigma_{B_2}^2}}
\label{eq:rho}
\end{eqnarray}
where the expected value and the variance of the Bernoulli variable $B_i$ is given by
\begin{eqnarray}
E[B_i]&=&p_i \label{eq:EBi} \\
\sigma_{B_i}^2&=&p_iq_i.\label{eq:sigmaBi}
\end{eqnarray}

By substituting  (\ref{eq:EBi}) and (\ref{eq:sigmaBi}) into (\ref{eq:rho}) and solving for $E[{B_1}{B_2}]$,
 \begin{eqnarray}
E[{B_1}{B_2}]=p_1p_2+\rho \sqrt{p_1p_2q_1q_2} = p_1p_2+\rho h.
\label{EB1B2}
\end{eqnarray}

We can relate $p_{B_1,B_2}(b_1,b_2)$ to $E[B_1B_2]$ as follows:
\begin{eqnarray}
E[B_1 B_2] &=& \sum_{b_1} \sum_{b_2} b_1 b_2 p_{B_1,B_2}(b_1,b_2) = p_{B_1,B_2}(1,1), \nonumber 
\end{eqnarray}
where solving the sum relies there being only one nonzero value for $b_1b_2$. 
By solving for $p_{B_1,B_2}(1,1)$ and using (\ref{EB1B2}),
\begin{eqnarray}
p_{B_1,B_2}(1,1)
&=& 
p_1 p_2+ \rho h.
\end{eqnarray}

We can relate $p_{B_1,B_2}(b_1,b_2)$ to $E[B_1]$ as follows:
\begin{eqnarray}
E[B_1] &=& \sum_{b_1} \sum_{b_2} b_1  p_{B_1,B_2}(b_1,b_2) \nonumber \\
&=& p_{B_1,B_2}(1,1)+p_{B_1,B_2}(1,0).
\end{eqnarray}

Solving for $p_{B_1,B_2}(1,0)$,
\begin{eqnarray}
p_{B_1,B_2}(1,0)&=& E[B_1]-p_{B_1,B_2}(1,1) = p_1q_2-\rho h. \nonumber
\end{eqnarray}

Similarly, it can be shown that
\begin{eqnarray}
p_{B_1,B_2}(0,1)= q_1p_2-\rho h.
\end{eqnarray}

Finally, since
\begin{eqnarray} 
\sum_{b_1} \sum_{b_2}p_{B_1,B_2}(b_1,b_2)
& = & 1,
\end{eqnarray}
it follows that
\begin{eqnarray}
p_{B_1,B_2}(0,0)\hspace{-0.3cm} &=&\hspace{-0.3cm} 1-p_{B_1,B_2}(1,0)-p_{B_1,B_2}(0,1)-p_{B_1,B_2}(1,1)\nonumber \\
& = &
q_1 q_2 + \rho h. 
\end{eqnarray}

\balance

\small
\bibliographystyle{ieeetr}

\balance
\bibliography{./References}


\end{document}